# A Context-aware Recommender System for Hyper-local News – A Conceptual Framework

Anh Nguyen Duc[1], Hilde Gudvangen[2]

[1]Department of Computer and Information Science, NTNU
[2]Muml AS
anhn@idi.ntnu.no, hilde@muml.community
Trondheim, Norway

**Abstract.** Recommender systems (RSs) have been popular in variety of application domains due to the increased demand for filtering and sorting items and information. Today, there is a numerous approaches and algorithms of data filtering and recommendations. This works presents a conceptual framework for constructing a mobile RS in hyper-local news domain. The mobile RS is designed to deal with specific requirements of news readers, such as spatial-temporal relevance, recency, real-time update and validated news. The implementation of the RS in a distributed file system is also discussed.

## 1 Introduction

In Big Data era, information has increased at an unprecedented rate and the information overload problem has become increasingly severe for online users. Nighty percent of all the data available today were created in the last two years [30]. In this context, RS plays an important role to bring meaningful and relevant information to individuals and business organizations [1]. Starting from mid 1990s, RSs became an independent research area with a large application domain, including e-commerce, multimedia, work and productivity, news, education and tourism [1-3]. This work investigates the feasibility of applying a RS to a hyper-local news mobile application.

The mission of a hyper-local news editor is to deliver relevant news to users as quick as possible, considering its location context. Hyper-local news is targeted at or consumed by people or entities that are located within a well-defined area, generally on the scale of a street, a neighborhood, a community or a city. Hyperlocal content must also be relevant in time. The higher the content scores on these dimensions the more relevant the content becomes to the individual and the less it becomes to the masses.

Mobile apps for hyper-local news are becoming popular in software startup scenes around the world, such as BlogFeed[1], Ripple[2] and MittMedia[3] to name a few. However, these startups are also facing with challenges of making sense out of the large volume of data occurring in a real-time manner. We are particularly interested in RSs for mobile application. Besides the mentioned concerns, mobile RSs face a challenge of making accurate recommendations using simple, yet appealing user interface [27]. Most of the mobile RSs heavily rely on locations of the users to recommend items to them, which is also essential in hyper-local news domain. Alternatively, the recommendation is made based on not only item's content but also user's context variables, i.e. geographical location and time.

This paper proposes a conceptual architecture of a mobile news RS applied for hyper-local news with user-generated contents and social network data. Our solution adopts different recommendation techniques and considers mobile-specific factors, such as geographical location and temporal information. We also discussed the proposed RS in big data perspective.

The paper is structured as follows: Section 2 presents background about recommendation approaches and challenges in recommending news. Section 3 describes requirements to a hyper-local news RS and Section 4 discuses its conceptual architecture. Finally, Section 5 concludes the paper with future perspectives.

## 2  Background

### 2.1. Recommendation techniques

State-of-the-art recommendation approaches can be summarized as in Table 1. Traditional recommendation approaches are classified into content based filtering, collaborative filtering and hybrid approaches [1,2]. Content-based filtering [6-8] utilizes several characteristics of a rated item to recommend a future item with similar characteristics. For example, a user selects a movie with a specific genre, IMDb score, and editor's evaluation. A content-based RS will probably recommend a movie with the same genre, IMDb score and editor's judgment for the user. Certain characteristics of an item, i.e. textual description, linked images or sound can be analyzed to find the similarity among items.

Collaborative filtering (CF) produces a recommendation using both user's past preference and also similar decisions made by other users [3-5]. The CF technique can be divided into user- based and item-based CF approaches [9]. In the user-based CF approach, a user will receive recommendations of items liked by similar users. In the item-based CF approach, a user will receive recommendations of items that are similar to those they were preferred in the past. Hybrid approaches combine multiple RS techniques to achieve a synergy between them. Several researchers have attempted to combine CF and content-based approaches in order to smoothen their disadvantages and to gain better performance [10-13].

---

[1] http://blockfeed.com

[2] https://ripple.co

[3] https://www.mittmedia.se

Recent advancement in RS considers additional characteristics of user reading [14-18]. Social-based RSs utilize social interactions among users, which are available in Internet, to improve the effectiveness of traditional recommendation mechanisms. The social interactions include online friending, making social comments, social tags, etc. Other types of social relations are also used for recommendation generation, i.e. social bookmarks, physical context, social tags, and "co-authorship" relations [14-18]. Particularly, trust based system puts a weight on the opinions of an user which is a friend or a person that the user can trust [14, 15].

Context-aware RSs utilize information such as time, geometrical information, or the company of other people (friends, families or colleagues for example), for some applications in which it is not sufficient to consider only users and items, such as recommending a vacation package, or personalized content on a website [2]. For example, using a temporal context, a travel recommender system might make a very different vacation recommendation in winter compared to summer [19]. The contextual information about users in technology enhanced learning environments is also incorporated into the recommendation process [20].

Group recommender system (GRS) applies for a group of users as a unit of analysis instead of an individual. While many RSs are focused on making recommendations to a single user, many daily activities such as watching a movie or going to a restaurant involve a group of users, in which case recommendations must take into account the tastes and preferences of all the users in the group [1].

**Table 1: Recommendation techniques**

| Domain | Techniques |
|---|---|
| Traditional | Content based filtering |
| | Collaborative filtering |
| | Hybrid approaches |
| More recent | Social based recommendation |
| | Context aware approach |
| | Group based recommender |

**2.2. The challenges of recommeding a hyper-local news**

News RS do not escape the common challenges with general RSs [21,26]. Özgöbek et al. discussed several issues in a news recommender system [21], as shown in Table 1. We added three news challenges that are specifically important for hyper-local news domain, namely context awareness, social concern and real-time update (notated by * in Table 2).

- Context awareness: in hyper-local news domain, readers would like to read news that are relevant to his location, i.e. what interesting things happen in a nearby street corner. When travelling, the information about traffic jam does only matter if it is on user's way.

- Social relation concern: readers would be more interested to read news from their friends or the ones they follow. This would require the integration of personal data from social networks.

- Real-time update: most of the traditional RSs have two phases, offline model construction and on-demand recommendation phase, where the model is fed with new data. The model can be updated at regular time intervals, e.g., hours or days, cannot meet the real-time demands [22, 23].

**Table 2: Challenges and requirements for RSs in hyper-local news domain**

| Problem | Description |
|---|---|
| Cold start | Little or no information about a new user/ item/ system when firstly introduced |
| Scalability | The ability of a RS to handle an ultra large set of data |
| Sparsity | insufficient various data leads to an user-item matrix with most of elements are zero |
| Gray sheep | it is not possible to recommend a proper item to a person whose preferences do not consistently agree or disagree with any group of people |
| Neighbour transivity | when dataset is very spare, two users with similar interest can not be detected due to the lack of their common ratings |
| Missing data | Data gathered from internet misses data field for generating recommendation, i.e. timestamp or news location |
| Privacy and security | RSs require access to private historical data of users |
| Serendipity | news written differently from different sources can be recommended as a different post |
| Recency | old news might be quickly obsolete and not interesting anymore to readers |
| Changing interest | the interest of reader changes over time. RSs become inaccurate until the system notices the change in the user interest. |
| Unstructured data | the news domain is characterized by fluctuating and unclear vocabularies and ever changing news topics. |
| Context awareness* | readers want to read news that are of their preferences and also fit to their current location and time |
| Social relation concern* | The decisions from persons who close to the user in social network, i.e. Facebook, Twitter, should be weighted higher. |

## 3. Requirements to a hyper-local news RS

Muml AS is a software startup located in Trondheim. The company develops a hyper-local news service to provide users with relevant and validated news in the real-time manner. Users who install the mobile app can be notified news that are of their interest happening nearby. The company has gone through initial startup phases by (1) refining the product ideas, (2) validating the market demand at different scales, (3) and building up the first Minimum viable product (MVP), a technically demonstrable version of the product. In a large city like Singapore, Kuala Lumpur and Hanoi, a vast various types of news, including shopping, events, traffic, neighborhood, restaurant and services were available for publish from every street corner. A preliminary market study on a local community[4] recorded thousands of post per hour. Big data issue is soon a challenge for scaling up the product.

Muml plans to embedd an advanced feature in the second MVP, so-called SmartGuide. The feature aims to provide intelligence for the service, by recommending news to an user in a real-time manner. Each piece of news is collected with the following attributes: (1) a photo/ a short video, (2) description, (3) category, (4) channel, (5) hashtag, (6) location, (7) time created and (8) user created. User's usage log is stored in a device and updated to a server frequently. The usage log includes: (1) read news list, (2) likes list and (3) comments.

Requirements for SmartGuide system has been set as below:
- The system should collect user's historical usage data from the user's device.
- Usage data should be associated with context information, i.e. user's location, timestamp.
- The system should, based on the usage data, recommend relevant news to the user.
- The relevance should consider factors: time, location, user's preference and user's friend's preference.
- The system should provide a real-time update.
- The system should deliver the recommended news queries, i.e., by users requests (pull-based).

## 4. Conceptual architecture

Muml is implemented as a standard mobile application with a client-sever architecture via REST. The frontend part is thin, presenting news to users in both a map view and a list view. The frontend has a cache to support store offline data and usage log. The backend implements all main logic functions, including the SmartGuide feature. The news is recommended to a user based on his reading history as well as other user's preference. Searching for relevant news in a greedy manner often impact on a long-term performance. We adopt a reinforcement learning approach to balance the tradeoff between exploration and exploitation. The main element of SmartGuide is shown as in Figure 1.

---

[4] beat.vn

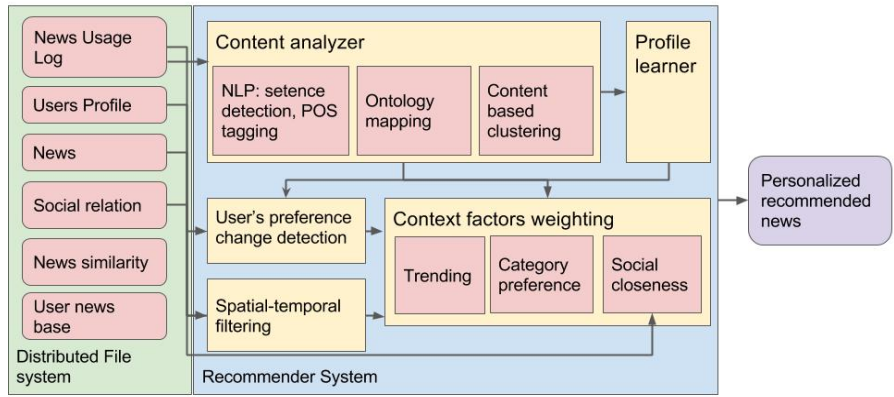

**Figure 1: Conceptual architecture of SmartGuide**

**Content analyzer component** analyzes the description part of a news, mapping the identified textual elements into predefined topics, as described in a previous research [28]. This information combines with other attributes (as in Section III) to provide a news profile.

**Profile learner component** updates a news profile database and a user profile database when exploring new data entering to the system. The user profile is initially empty. Based on a reinforcement algorithm [31], when user select a news to read, the user profile is updated, either by adding new preferences or by updating reward/punishment values associated to existing elements of the profile. The specific algorithm in charge of managing these punishments/rewards is Q-learning [32]. The core part is a storage of state-action pairs and we can learn from the changing of value state $Q(s,a)$ between state- action pair to another state-action pair. An abstract news content can be viewed as states s of the system and moving can be viewed as action a of state:

$$Q(s,a) = Q(s,a) + \alpha[r_{t+1} + \gamma Q(s_{t+1}, a_{t+1}) - Q(s_t, a_t)]$$

**Change detection component** calculates two types of preference for each user, short-term and long-term preference. The change is detected when there is a derivation of short-term interest's value [29]. Particularly, more weight is given to a news category that has drawn recent user interest.

**Spatial-temporal filtering component** limits the exploration domain by excluding news that are out of the interest scope, i.e. news that are outside 5 km radius and beyond 24 hours. The current configuration is predetermined for a metropolitan context. It is can also configured for other context setting.

**Context factors weighting component** incorporates the influences of contextual factors, so that the trendy news will be weighted more, the news read by user's friends will be weighted more and the news are of user's preference will be weighted more. The framework is designed to solve the recency issue by giving more weights to the more recent news. We also consider diversity of news by adjusting weights given alongside trends, categories and social relations.

**Perspective on Big data** in our architecture is the deployment of data storage and query in distributed file system platforms. i.e. Hadoop and Spark. Three elements will be parallel processed in the distributed file systems: (1) Content analyzer, (2) Profile learner and (3) Context factor weighting. Data processing will eventually end in storing two tables: New_Similarity (News_ID, Similar_News, Similarity_Score) and User_News_Base (User_ID, News_ID, Recommendation_Score). The implementation of the architecture is split into an offline process and online process. The data preprocessing and analysis of news usage log, user profile, news database, and social relation is done in batch in a daily/ weekly basis. The online process composes querying new similarity and user news base tables. Online data processing includes the handling of the weekly results and also the most recent updated user read log.

## 6. Discussion and Conclusion

Without RSs, all users might read the same set of news. RSs help to filter relevant news to specific users, given his reading profiles or other personal information; and to ensure audiences for all types of news, even for a niche area, i.e. startup and innovation. Under the era of big data, new development of RS techniques are required to deal with the evolving data volume as well as requirements of real-time updates. This paper describes a solution towards a real-time mobile RS for a hyper-local news application. Building on top of existing work, we provide a new angle of RS research by introducing a simplifying mechanism to RS dataset, considering important issues of RS, such as recency, unstructured data, social relation concern, user context awareness and real-timeness. We discussed the implementation and deployment of SmartGuide in distributed infrastructure.

This study represents an on-going software development project. The next step would be to validate and refine the RS model and implement it in a real-life context. Derived from actual requirement of the project, future work can consider an enhancing version of RSs to support:

- News and scheduling services: integration the news RS mechanism to support user-scheduling his/ her daily life activity. For example, based on the information about traffic, the system can recommend users to take another road to home at the end of his workday.
- Privacy and security: recommended items largely depend on stored user profiles, which hold privacy-sensitive information. In the future, a tailored mobile RS methodology for protecting user anonymity and privacy are desirable.